\documentclass[12pt,aps,prd,showpacs,amsmath,amssymb]{revtex4}
\usepackage{amssymb}
\input epsf
\allowdisplaybreaks
\begin{document}
\title{The exclusive rare decay $b\rightarrow s\gamma$ of heavy b-Baryons}
\author{Yong-Lu Liu}
\author{Long-Fei Gan}
\author{Ming-Qiu Huang}
\affiliation{Department of Physics, College of Science, National University of Defense Technology, Changsha, Hunan 410073, People's Republic of China}
\date{\today}
\begin{abstract}
We present an analysis on the exclusive rare radiative decay modes $\Sigma_b\rightarrow\Sigma\gamma$ and $\Xi_b\rightarrow\Xi\gamma$. The transition form factors which parameterize these processes are calculated using QCD light-cone sum rules. The decay widths we predict are $\Gamma(\Sigma_b\rightarrow\Sigma\gamma)=(7.21\pm0.04)\times10^{-18}\,\mbox{GeV}$ and $\Gamma(\Xi_b\rightarrow\Xi\gamma)=(1.34\pm0.07)\times10^{-16}\,\mbox{GeV}$. The Branching ratio of $\Xi_b\rightarrow\Xi\gamma$ is predicted to be $\mbox{Br}(\Xi_b\rightarrow\Xi\gamma)=(3.03\pm0.10)\times10^{-4}$.
\end{abstract}
 \maketitle

\section{Introduction}
\label{sec1}
Heavy flavor physics plays an important role both in the precise test of the standard model in the relatively high energy region and in the investigation of the hadronization of quarks at the low energy. Hence, a lot of effort has been paid into it and a large amount of experimental data have been accumulated \cite{Acosta1,Abazov,Aaltonen,Aaltonen2}. Theoretically, much progress has been made in the heavy flavor meson sector for its comparatively simple structure while knowledge about baryons appears to be limited. Although much literature have been provided to decipher these heavy flavor states (such as Refs. \cite{Ebert,Zhang,Roncaglia,Mathur}), a deep understanding of them undoubtedly demands the information on the dynamical details which are encoded in various decay modes \cite{Azizi,Huang}. Among these modes, the rare radiative decay processes of the b-baryons are important in that they are not only the ways to study the Cabibbo-Kobayashi-Maskawa matrix elements $V_{ts}$
and $V_{tb}$ which are closely attached with the dynamics inside the baryons, but also the ways to probe new physics beyond the standard model.

These types of processes (such as $b\rightarrow s\gamma$), which are forbidden at the tree level in the standard model of electroweak theory, are induced by the flavor-changing neutral current (FCNC) of the $b$-quark. Their amplitudes are dominated by the one-loop diagrams with a virtual top quark and a W boson, and thus are strongly suppressed by Glashow-Iliopoulos-Maiani mechanism. The relative b-meson rare radiative decay modes have been investigated experimentally since the early 1990's \cite{CLEO1,CLEO,BABAR,BELLE}, while not so many experimental data are available for the corresponding b-baryon processes. Theoretical studies on the exclusive processes are available for both b-mesons and b-baryons \cite{Mannel,Mohanta,HaiYang,Chun,hurth}, despite the fact that the dynamics of the b-baryons decays are far less clarified in comparison with that of the b-mesons. However, most of the existing literature is about the process $\Lambda_b\rightarrow\Lambda\gamma$, the branching ratio of which has been predicted to be $Br(\Lambda_b\rightarrow\Lambda\gamma)\le 1.3\times10^{-3}$ experimentally \cite{Acosta}. Unfortunately, this decay mode is not expected to be measured easily in the experiments due to the fact that the final state $\Lambda$ baryon is of neutral charged, as argued in Ref. \cite{Olive}. For this reason, we turn to study the possible decay modes of other octet heavy baryons $\Sigma_b$ and $\Xi_b$, in which charged final states arise and may be easily tested in experiments. It has been estimated early in the $1990's$ that an amount number of b-baryons may be produced at the c.m. energy level of the LHC \cite{Fridman}. Thus we can expect that these rare decay modes could be measured by the LHC experiments in the near future, the updated energy of which is expected to be $\sim 14\,\mbox{TeV}$.

The remainder of this paper is organized as follows. We give an introduction to the exclusive rare decay mode $b\rightarrow s\gamma$ and derive the formula of the decay widths in Sec. \ref{sec:2}.  Then the light-cone QCD sum rules for the relative transition form factors are derived in Sec. \ref{sec:3}. Finally, Sec. \ref{sec:4} is devoted to the numerical analysis and a summary is given at the end of this section.

\section{Parametrization of the transition form factors}\label{sec:2}
In the standard model, the process of the exclusive rare decay $b\rightarrow s\gamma$ can be described by the following effective Hamiltonian \cite{Buchalla}:
\begin{equation}
H_{eff}(b\rightarrow s\gamma) = -4\frac{G_F}{\sqrt{2}}  V_{t s}^\ast  V_{tb}
  C_{7}(\mu)  {\cal O}_7(\mu),
\end{equation}
with
\begin{equation}
{\cal O}_{7} = \frac{e }{ 16\ \pi^2}\bar{s}\ \sigma_{\mu \nu}\
(m_{b} R+ m_{s} L)\ b\ F^{\mu \nu},
\end{equation}
where $L/R=(1\mp\gamma_5)/2$ and $F^{\mu \nu}$ is the field strength
tensor of the photon. $G_F$ is the Fermi coupling constant and
$C_{7}(\mu)$ is the Wilson coefficient at the scale $\mu$. Considering the general form beyond the standard model, ${\cal O}_{7}$ can be represented as
\begin{equation}
{\cal O}_{7} = \frac{e }{ 32\ \pi^2}m_b\bar{s}\ \sigma_{\mu \nu}\
(g_V+\gamma_5 g_A)\ b\ F^{\mu \nu}.
\end{equation}

The decay amplitude is given by the expectation value of the effective Hamiltonian
between the initial and final states at the hadron level
\begin{equation}
M(X_b\rightarrow X\gamma)=\langle
X\gamma|H_{eff}|X_b\rangle\;,
\end{equation}
where $X$ stands for the baryon involved in the process.

By considering the Lorentz structure, the contribution of the hadronic part to the process, which is written as the hadronic matrix elements, is generally parameterized in terms of the following form factors:
\begin{equation}\label{matr}
\langle X_b(P')|j_\nu|X(P)\rangle=\bar X_b(P')[f_1\gamma_\nu-f_2i\sigma_{\mu\nu}q^\mu+f_3q_\nu-(g_1\gamma_\nu+g_2i\sigma_{\mu\nu}q^\mu+g_3q_\nu)\gamma_5]X(P),
\end{equation}
where $X_b$ and $X$ are the spinors of the baryons and the weak current $j_\nu$ is defined as
\begin{equation}
j_\nu(x)=i\bar b(x)\sigma_{\mu\nu}(1-\gamma_5)q^\mu s(x).
\end{equation}
In fact, form factors $f_3$ and $g_3$ do not contribute to the process due to the conservation of the vector current. Therefore, the form factors we need to calculate are $f_1(g_1)$ and $f_2(g_2)$, which can be determined from the QCD light-cone sum rules. It is noted that the processes are only related to the form factors at the point $q^2=0$, thus we just consider this case in the following analysis.

With the form factors defined above, the decay width is represented as
\begin{equation}
\Gamma(X_b\rightarrow X\gamma)=\frac{G_F^2|V_{tb}V_{ts}^*|^2
\alpha_{em}|C_7|^2 m_b^2}{32\pi^4}\left(
\frac{M_{X_b}^2-M_X^2}{M_{X_b}} \right )^3(g_V^2
f_2^2 + g_A^2 g_2^2). \label{decayrate}
\end{equation}

\section{light-cone sum rules for the form factors}\label{sec:3}
Now we apply the light-cone QCD sum rule approach to calculate the transition form factors $f_1(g_1)$ and $f_2(g_2)$. The interpolating currents to the heavy baryons are chosen as $j_{\Sigma_b}(0)=\epsilon^{ijk}[q^i(0)C\rlap/zq^j(0)]\gamma_5\rlap/zb^k(0)$ for $\Sigma_b$ and $j_{\Xi_b}(0)=\epsilon^{ijk}[s^i(0)C\rlap/zb^j(0)]\gamma_5\rlap/zq^k(0)$ for $\Xi_b$, respectively. Herein $q$ stands for $u$ or $d$ quark, $C$ is the charge conjugaion matrix, and $z$ is the vector defined on the light-cone $z^2=0$. The normalization of these currents is defined by the parameters $f_{\Sigma_b}$ and $f_{\Xi_b}$:
\begin{eqnarray}\label{norm}
\langle0|j_{\Sigma_b}|\Sigma_b(P')\rangle&=&f_{\Sigma_b}(z\cdot P')\rlap/z\Sigma_b(P'),\nonumber\\
\langle0|j_{\Xi_b}|\Xi_b(P')\rangle&=&f_{\Xi_b}(z\cdot P')\rlap/z\Xi_b(P').
\end{eqnarray}

In the following part, we will take $\Sigma_b^+\rightarrow\Sigma^+\gamma$ as an example. Our starting point for calculating the form factors is the correlation function
\begin{equation}
T_\mu=i \int d^4x e^{iqx}\langle 0|{j_{\Sigma_b(0)}j_\nu(x)}|\Sigma(P,s)\rangle  \label{corr}
\end{equation}
at $q^2=0$ and with Euclidean $m_b^2-{P^\prime}^2$ of about several $\mbox{GeV}^2$.
Following the standard procedure of the light-cone sum rule method, we need to express the correlation function both phenomenologically and theoretically. By inserting a complete set of intermediate states and using the definitions (\ref{matr}) and (\ref{norm}), the phenomenological side is represented as
\begin{equation}
z^\nu T_\nu(P,q)=\frac{2f_{\Sigma_b}(z\cdot P')^2}{M_{\Sigma_b}^2-P'^2}[f_1\rlap/z-f_2\rlap/z\rlap/q-g_1\rlap/z\gamma_5+g_2\rlap/z\rlap/q\gamma_5]\Sigma(P)+...,
\end{equation}
where ``$...$" stands for the continuum contributions. The correlation function (\ref{corr}) is contracted by $z^\nu$ to remove contributions proportional to the light-cone vector $z^\nu$ which is subdominant on the light-cone.

On the other hand, the theoretical side is obtained by contracting the heavy $b$ quarks in the correlation function and using the distribution amplitudes presented in Ref.\cite{Braun,das-sig,das-xi}. To make the paper self-contained, we present in the Appendix the definition and the explicit expressions of the distribution amplitudes of $\Sigma$ and $\Xi$ used in this paper. After assuming the quark-hadron duality and performing the Borel transformation, we arrive at the final light-cone sum rule of the form factor $f_2(0)$:
\begin{eqnarray}\label{sumrule}
f_{\Sigma_b}f_2(0)e^{-\frac{M_{\Sigma_b}^2}{M_B^2}}&=&\int_{\alpha_{30}}^1d\alpha_3e^{-\frac{s}{M_B^2}}
\Big\{B_0(\alpha_3)+\frac{M^2}{M_B^2}B_1(\alpha_3)-\frac{M^4}{M_B^4}B_2(\alpha_3)\Big\}\nonumber\\
&&-\frac{M^2\alpha_{30}^2e^{-\frac{s_0}{M_B^2}}}{\alpha_{30}^2M^2+m_b^2}\Big\{B_1(\alpha_{30})
-\frac{M^2}{M_B^2}B_2(\alpha_{30})-\frac{d}{d\alpha_{30}}\frac{\alpha_{30}^2M^2B_2(\alpha_{30})}{\alpha_{30}^2M^2+m_b^2}\Big\},
\end{eqnarray}
where $s=(1-\alpha_3)M^2+m_b^2/\alpha_3$, $M$ is the mass of the final baryon, and $M_B^2$ is the Borel parameter which is introduced to suppress the contributions from the higher resonances and the continuum states. Our calculation shows that $f_1=g_1=0$ and $f_2=g_2$. In Eq. (\ref{sumrule}), the following abbreviations are used for convenience:
\begin{eqnarray}
B_0(\alpha_3)&=&\int_0^{1-\alpha_3}d\alpha_1V_1(\alpha_1,1-\alpha_1-\alpha_3,\alpha_3),\nonumber\\
B_1(\alpha_3)&=&(2\widetilde V_1-\widetilde V_2-\widetilde V_3-\widetilde V_4-\widetilde V_5)(\alpha_3),\nonumber\\
B_2(\alpha_3)&=&(-\widetilde{\widetilde V_1}+\widetilde{\widetilde V_2}+\widetilde{\widetilde V_3}+\widetilde{\widetilde V_4}+\widetilde{\widetilde V_5}-\widetilde{\widetilde V_6})(\alpha_3).
\end{eqnarray}

The distribution amplitudes with tildes which come from the
integration by parts in $\alpha_3$ are defined as
\begin{eqnarray}\label{tilde1}
\widetilde
V_i(\alpha_3)&=&\int_0^{\alpha_3}d{\alpha_3'}\int_0^{1-\alpha_3'}d\alpha_1
V_i(\alpha_1,1-\alpha_1-\alpha_3',\alpha_3'), \nonumber\\
\widetilde{\widetilde
V_i}(\alpha_3)&=&\int_0^{\alpha_3}d{\alpha_3'}\int_0^{\alpha_3'}
d{\alpha_3''}
\int_0^{1-{\alpha_3''}}d\alpha_2V_i(
\alpha_1,1-\alpha_1-\alpha_3'',\alpha_3'').
\end{eqnarray}

The same procedure is also carried out to calculate the transition form factors of the process $\Xi_b\rightarrow \Xi\gamma$. We obtain the final sum rule as follows:
\begin{eqnarray}\label{sumrule2}
f_{\Xi_b}f_2(0)e^{-\frac{M_{\Xi_b}^2}{M_B^2}}&=&\int_{\alpha_{20}}^1d\alpha_2e^{-\frac{s'}{M_B^2}}
\Big\{C_0(\alpha_2)+\frac{M^2}{M_B^2}C_1(\alpha_2)+\frac{M^2}{\alpha_2M_B^2}C_2(\alpha_2)-\frac{M^4}{M_B^4}C_3(\alpha_2)\Big\}\nonumber\\
&&+\frac{M^2\alpha_{20}^2e^{-\frac{s_0}{M_B^2}}}{\alpha_{20}^2M^2+m_b^2}\Big\{C_1(\alpha_{20})+\frac{1}{\alpha_{20}}C_2(\alpha_{20})
-\frac{M^2}{M_B^2}C_3(\alpha_{20})\nonumber\\
&&+\frac{d}{d\alpha_{20}}\frac{\alpha_{20}^2M^2C_3(\alpha_{20})}{\alpha_{20}^2M^2+m_b^2}\Big\},
\end{eqnarray}
where $s'=(1-\alpha_2)M^2+m_b^2/\alpha_2$, $M$ is the mass of $\Xi$, and the following abbreviations are used:
\begin{eqnarray}
C_0(\alpha_2)&=&\int_0^{1-\alpha_2}d\alpha_1T_1(\alpha_1,\alpha_2,1-\alpha_1-\alpha_2),\nonumber\\
C_1(\alpha_2)&=&(2\widetilde T_1-\widetilde T_2-\widetilde T_5-2\widetilde T_7-2\widetilde T_8)(\alpha_2),\nonumber\\
C_2(\alpha_2)&=&(\widetilde{\widetilde T_2}-\widetilde{\widetilde T_3}-\widetilde{\widetilde T_4}+\widetilde{\widetilde T_5}+\widetilde{\widetilde T_7}+\widetilde{\widetilde T_8})(\alpha_2),\nonumber\\
C_3(\alpha_2)&=&(-\widetilde{\widetilde T_1}+\widetilde{\widetilde T_2}+\widetilde{\widetilde T_5}-\widetilde{\widetilde T_6}+2\widetilde{\widetilde T_7}+2\widetilde{\widetilde T_8})(\alpha_2).
\end{eqnarray}

The functions with tildes are defined as
\begin{eqnarray}
\widetilde
T_i(\alpha_2)&=&\int_0^{\alpha_2}d{\alpha_2'}\int_0^{1-\alpha_2'}d\alpha_1
T_i(\alpha_1,\alpha_2',1-\alpha_1-\alpha_2'), \nonumber\\
\widetilde{\widetilde
T_i}(\alpha_2)&=&\int_0^{\alpha_2}d{\alpha_2'}\int_0^{\alpha_2'}
d{\alpha_2''}
\int_0^{1-{\alpha_2''}}d\alpha_2T_i(
\alpha_1,\alpha_2'',1-\alpha_1-\alpha_2''). \label{tilde}
\end{eqnarray}

\section{numerical analysis and the summary}\label{sec:4}
Before the numerical evaluation of the sum rules (\ref{sumrule}) and (\ref{sumrule2}), we need to determine the input parameters. Two important parameters are the decay constants $f_{\Sigma_b}$ and $f_{\Xi_b}$, which can be calculated with the QCD sum rule approach. Using the same expressions in Refs. \cite{das-sig} and \cite{das-xi} with the replacements $m_s\rightarrow m_b$ for $f_{\Sigma_b}$ and $m_c\rightarrow m_b$ for $f_{\Xi_b}$, we get the estimations $f_{\Sigma_b}=(6.18\pm0.03)\times10^{-3}\,\mbox{GeV}^2$ and $f_{\Xi_b}=(3.32\pm0.46)\times10^{-3}\,\mbox{GeV}^2$.
Other input parameters needed in our calculation can be read from Ref. \cite{PDG}:
\begin{eqnarray}
m_b&=&4.8\,\mbox{GeV},\,m_s=0.15\,\mbox{GeV},\,M_\Sigma=1.189\,\mbox{GeV},\nonumber\\
M_\Xi&=&1.314\,\mbox{GeV},\,M_{\Sigma_b}=5.729\,\mbox{GeV},\,M_{\Xi_b}=5.81\,\mbox{GeV},\nonumber\\
V_{ts}&=&0.0403,V_{tb}=0.9992,
\end{eqnarray}
and
\begin{eqnarray}
\alpha_{em}=1/137,G_F=1.166364\times10^{-5}\,\mbox{GeV}^{-2},C_7(m_b)=-0.31.
\end{eqnarray}

An important step in the numerical analysis of the QCD sum rules is to determine the Borel mass parameter $M_B^2$ and the continuum threshold $s_0$. The continuum threshold $s_0$ can be chosen by demanding that the continuum contribution is subdominant in comparison with that of the ground state which we are concerned about. Simultaneously, the resulting form factors should not vary drastically along with the threshold. Thus $s_0$ is generally connected with the first resonance which has the same quantum numbers as the particle we care about. Here we fix the threshold $s_0$ in the region $39\,\mbox{GeV}^2\le s_0\le 41\,\mbox{GeV}^2$. As for the Borel parameter $M_B^2$, which is introduced to suppress the higher resonance contributions efficiently, we also demand that the higher twists contributions are less significant and the form factors should vary mildly along with it. Our calculation shows that the working windows can be chosen properly in the region $8\,\mbox{GeV}^2\le M_B^2\le 11\,\mbox{GeV}^2$ for $\Sigma_b\rightarrow\Sigma\gamma$ and $9\,\mbox{GeV}^2\le M_B^2\le 12\,\mbox{GeV}^2$ for $\Xi_b\rightarrow\Xi\gamma$.
\begin{figure}
\begin{minipage}{7cm}
\epsfxsize=7cm \centerline{\epsffile{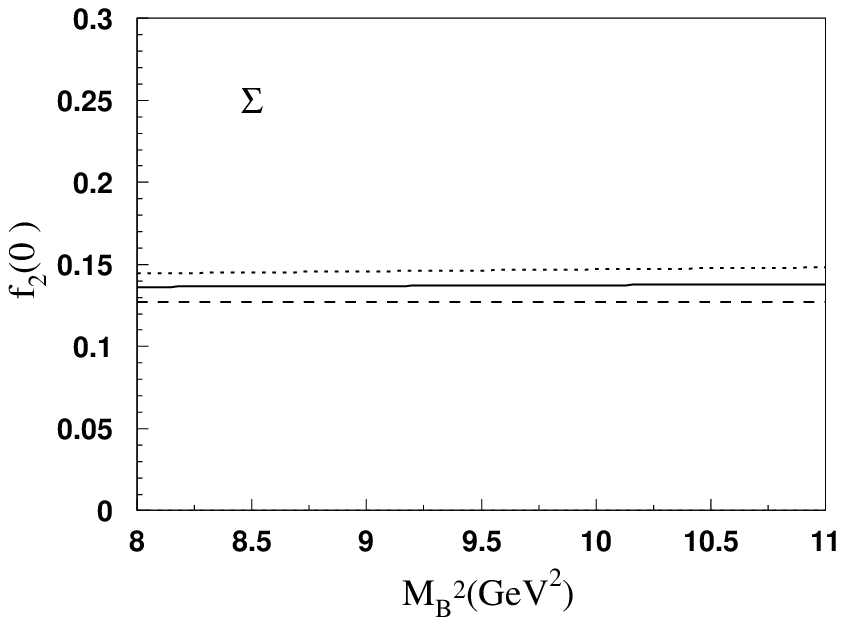}}
\end{minipage}
\begin{minipage}{7cm}
\epsfxsize=7cm \centerline{\epsffile{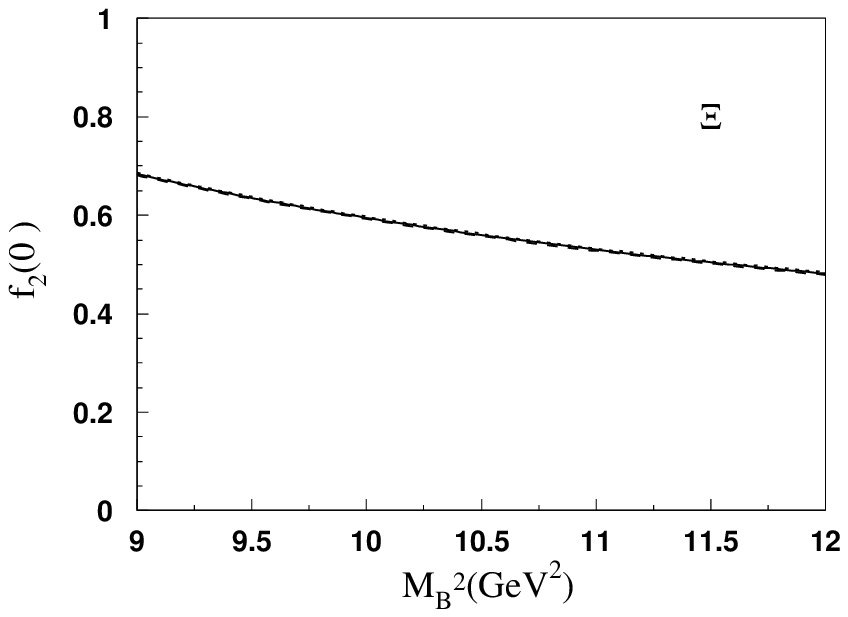}}
\end{minipage}
\caption{\quad The dependence of the form factors $f_2(0)$'s on the Borel parameter with $s_0=39,\,40,\,41\,\mbox{GeV}^2$ from the top down. }\label{fig1}
\end{figure}

\begin{figure}
\begin{minipage}{7cm}
\epsfxsize=7cm \centerline{\epsffile{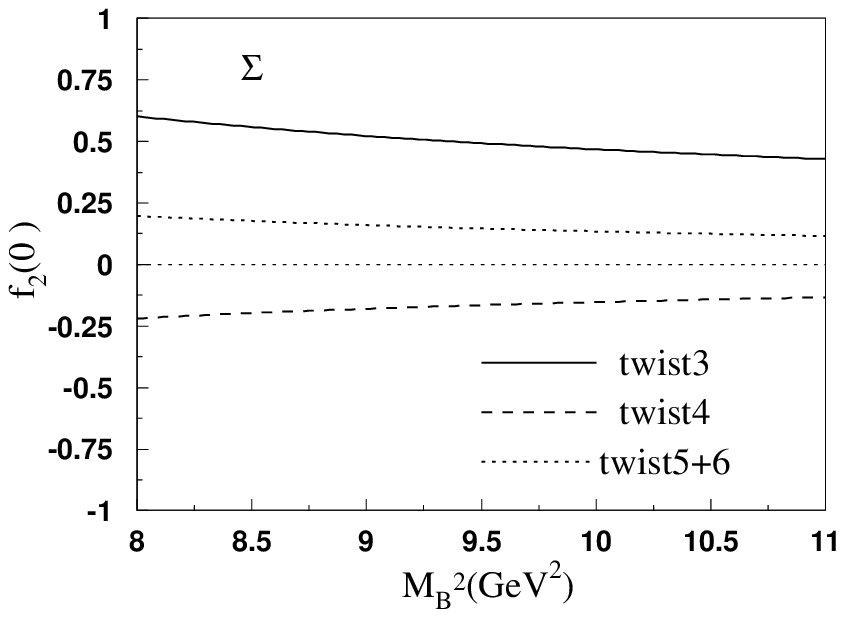}}
\end{minipage}
\begin{minipage}{7cm}
\epsfxsize=7cm \centerline{\epsffile{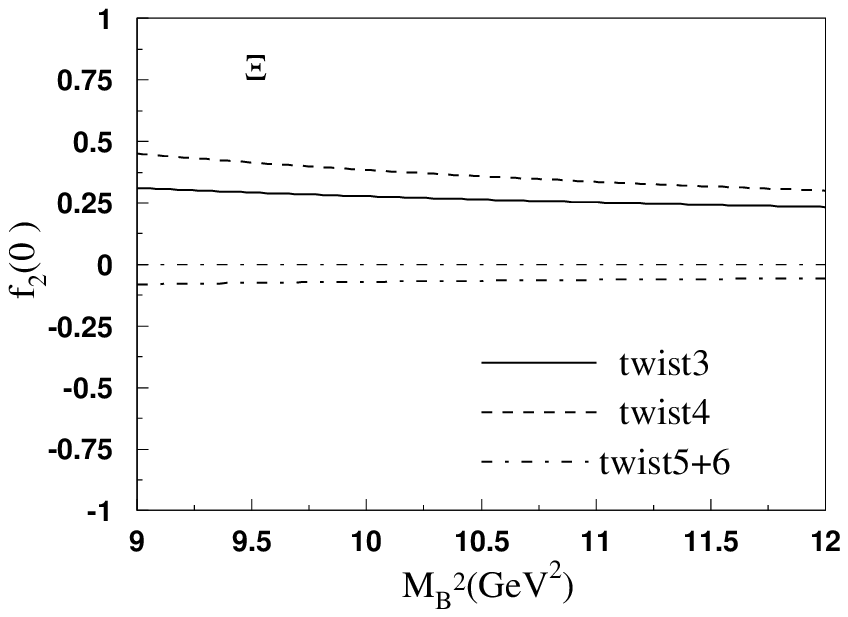}}
\end{minipage}
\caption{\quad The contributions to the form factors $f_2(0)$'s from different twists on the Borel parameter with $s_0=40\,\mbox{GeV}^2$. }\label{fig2}
\end{figure}

Using the distribution amplitudes given in Refs. \cite{das-sig} and \cite{das-xi}, we obtain the form factors at the zero momentum transfer $f_2(g_2)(0)$ as functions of the Borel parameter $M_B^2$, which are displayed in Fig. \ref{fig1} . We have also analyzed the contributions from the distributions of different twists, which are shown in Fig. \ref{fig2}. The results show that the contributions of the leading-order and next-to-leading-order twists are dominant while the contributions from higher twists are suppressed efficiently. This implies that the light-cone expansion is reasonable in the cases we considered in this paper.

By using of the form factors we have estimated above, the decay widths of the processes can be easily evaluated with the formula (\ref{decayrate}), which turn out to be $\Gamma(\Sigma_b\rightarrow\Sigma\gamma)=(7.21\pm0.04)\times10^{-18}\,\mbox{GeV}$ and $\Gamma(\Xi_b\rightarrow\Xi\gamma)=(1.34\pm0.07)\times10^{-16}\,\mbox{GeV}$. Although the mean life of $\Xi_b^-$ has been estimated experimentally \cite{Aaltonen,Aaltonen2,Buskulic,Abdallah}, here we use the average value given in Ref. \cite{PDG} to estimate the branching ratio of the process $\Xi_b^-\rightarrow\Xi^-\gamma$ which turns out to be $Br(\Xi_b^-\rightarrow\Xi^-\gamma)=(3.03\pm0.10)\times10^{-4}$. The errors in the widths come from the choices of the threshold, the sum rule windows, and the uncertainties in the decay constants $f_{\Sigma_b}$ and $f_{\Xi_b}$. It is worth noting that errors from other sources are not considered here because the sum rule method itself brings in an amount of uncertainties (about $20\%$), which makes it less significant to take into account the errors of the input parameters.

We also investigate the sensitivity of the form factors to the variation of $m_b$ at different points $m_b=4.7,\,4.8,\,$ and $4.9\,\mbox{GeV}$. The corresponding predictions for the decay widths and branching ratios are given in Table \ref{table}.
\begin{table}
\caption{Decay widths and Branching ratios at different points of $m_b$ .}
\begin{center}
\begin{tabular}
{|c|c|c|c|}
\hline  $m_b(\mbox{GeV})$ & $4.7$ & $4.8$ & $4.9$  \\
\hline  $\Gamma(\Sigma_b\rightarrow\Sigma\gamma)(\times10^{-18}\,\mbox{GeV})$ & $6.92\pm0.03$ & $7.21\pm0.04$ & $7.26\pm0.07$  \\
\hline  $\Gamma(\Xi_b\rightarrow\Xi\gamma)(\times10^{-16}\,\mbox{GeV})$ & $0.98\pm0.04$ &$1.34\pm0.04$ & $1.75\pm0.05$  \\
\hline  $Br(\Xi_b^-\rightarrow\Xi^-\gamma)(\times10^{-4})$ & $2.21\pm0.08$ & $3.03\pm0.10$ & $3.96\pm0.11$  \\
\hline
\end{tabular}
\end{center} \label{table}
\end{table}

In summary, we have investigated the exclusive rare decay processes $\Sigma_b\rightarrow \Sigma\gamma$ and $\Xi_b\rightarrow\Xi\gamma$. The corresponding transition form factors are estimated through the light-cone QCD sum rule approach and the decay widths of these processes are predicted to be $\Gamma(\Sigma_b\rightarrow\Sigma\gamma)=(7.21\pm0.04)\times10^{-18}\,\mbox{GeV}$ and $\Gamma(\Xi_b\rightarrow\Xi\gamma)=(1.34\pm0.04)\times10^{-16}\,\mbox{GeV}$. We also estimate the branching ratio of $\Xi_b^-\rightarrow\Xi^-\gamma$, which is $Br(\Xi_b^-\rightarrow\Xi^-\gamma)=(3.03\pm0.10)\times10^{-4}$. As we can see, our prediction is larger than the theoretical estimations for the branching ratio of the $\Lambda_b\rightarrow\Lambda\gamma$. Therefore it is reasonable to assume that this mode may be tested easily, provided that a good source of $\Sigma_b$ or $\Xi_b$ is available in the future experiments, such as the LHC experiments.

%%%%%%%%%%%%%%%%%%%%%%%%%%%%%%%%%%%%%%%%%%%%%%%%%%%%%%%%%
\acknowledgments  This work was supported in part by the National
Natural Science Foundation of China under Contract Nos.10975184 and 11047117.
%%%%%%%%%%%%%%%%%%%%%%%%%%%%%%%%%%%%%%%%%%%%%%%%%%%%%%%%
\appendix
%%%%%%%%%%%%%%%%%%%%%%%%%%%%%%%%%%%%%%%%%%%%%%%%%%%%%%%%%%%%%%%%%%%%%%%%%%%%%
\section*{Appendix}
In the following we give the distribution amplitudes of $\Sigma$ and $\Xi$ used in the paper. In general, the distribution amplitudes are defined by the matrix element of the three-quark operator as
\begin{eqnarray}
&& 4\langle {0}| \epsilon^{ijk} {q_1}_\alpha^i(a_1 z) {q_2}_\beta^j(a_2 z) {q_3}_\gamma^k(a_3 z) |{X(P)} \rangle
\nonumber \\
&=& V_1  \left(\!\not\!{p}C \right)_{\alpha \beta} \left(\gamma_5 X^+\right)_\gamma + V_2  \left(\!\not\!{p}C \right)_{\alpha \beta} \left(\gamma_5
X^-\right)_\gamma + \frac{V_3}{2} M \left(\gamma_\perp C \right)_{\alpha \beta}\left( \gamma^{\perp} \gamma_5 X^+\right)_\gamma
\nonumber \\
&& + \frac{V_4}{2} M  \left(\gamma_\perp C \right)_{\alpha \beta}\left( \gamma^{\perp} \gamma_5 X^-\right)_\gamma + V_5 \frac{M^2}{2 p z} \left(\!\not\!{z}C
\right)_{\alpha \beta} \left(\gamma_5 X^+\right)_\gamma + \frac{M^2}{2 pz} V_6 \left(\!\not\!{z}C \right)_{\alpha \beta} \left(\gamma_5 X^-\right)_\gamma
\nonumber \\
&& + T_1 \left(i \sigma_{\perp p} C\right)_{\alpha \beta} \left(\gamma^\perp\gamma_5 X^+\right)_\gamma + T_2 \left(i \sigma_{\perp\, p} C\right)_{\alpha \beta}
\left(\gamma^\perp\gamma_5 X^-\right)_\gamma\nonumber \\
&& + T_3 \frac{M}{p z} \left(i \sigma_{p\, z} C\right)_{\alpha \beta} \left(\gamma_5 X^+\right)_\gamma + T_4 \frac{M}{p z}\left(i \sigma_{z\, p} C\right)_{\alpha
\beta} \left(\gamma_5 X^-\right)_\gamma \nonumber \\
&&+ T_5 \frac{M^2}{2 p z}  \left(i \sigma_{\perp\, z} C\right)_{\alpha \beta} \left(\gamma^\perp\gamma_5 X^+\right)_\gamma + \frac{M^2}{2 pz}  T_6 \left(i
\sigma_{\perp\, z} C\right)_{\alpha \beta} \left(\gamma^\perp\gamma_5 X^-\right)_\gamma \nonumber\\ && + M \frac{T_7}{2} \left(\sigma_{\perp\, \perp'}
C\right)_{\alpha \beta} \left(\sigma^{\perp\, \perp'} \gamma_5 X^+\right)_\gamma+ M \frac{T_8}{2} \left(\sigma_{\perp\, \perp'} C\right)_{\alpha \beta}
\left(\sigma^{\perp\, \perp'} \gamma_5 X^-\right)_\gamma\,,\label{da-deftwist}
\end{eqnarray}
where $M$ is the mass of the baryon $X$ and $C$ is the charge conjugation matrix. Note that the other Lorentz structures which do not contribute to the calculations are omitted. For each distribution amplitudes $F_i=V_i\,,T_i$ defined above, it can be presented as
\begin{equation}
F(a_ip\cdot z)=\int \mathcal Dxe^{-ipz\sum\limits_ix_ia_i}F(x_i)\,,
\end{equation}
with the relationship $0<x_i<1,\; \sum\limits_ix_i=1$, and $x_i$ corresponds to the distribution of the baryon momentum on the quarks. The integration measure is defined as
\begin{equation}
\int \mathcal Dx=\int_0^1dx_1dx_2dx_3\delta(x_1+x_2+x_3-1)\,.
\end{equation}

The distribution amplitudes can be expanded with a conformal spin. The detailed process is referred to Refs. \cite{Braun,das-sig,das-xi}. The explicit expressions of the distribution amplitudes are collected below:
\begin{eqnarray}
V_1(x_i)&=&120x_1x_2x_3\phi_3^0\,,\hspace{2.4cm}V_2(x_i)=24x_1x_2\phi_4^0\,,\nonumber\\
V_3(x_i)&=&12x_3(1-x_3)\psi_4^0\,,\hspace{1.9cm}V_4(x_i)=3(1-x_3)\psi_5^0\,,\nonumber\\
V_5(x_i)&=&6x_3\phi_5^0\,,\hspace{3.6cm}V_6(x_i)=2\phi_6^0\,,\nonumber\\
T_1(x_i)&=&120x_1x_2x_3\phi_3^{'0}\,,\hspace{2.3cm}T_2(x_i)=24x_1x_2\phi_4^{'0}\,,\nonumber\\
T_3(x_i)&=&6x_3(1-x_3)(\xi_4^0+\xi_4^{'0})\,,\hspace{1cm}T_4(x_i)=-\frac32(x_1+x_2)(\xi_5^{'0}+\xi_5^0)\nonumber\\
T_5(x_i)&=&6x_3\phi_5^{'0}\,,\hspace{3.6cm}T_6(x_i)=2\phi_6^{'0}\nonumber\\
T_7(x_i)&=&6x_3(1-x_3)(\xi_4^{'0}-\xi_4^0)\,,\hspace{1cm}T_8(x_i)=\frac32(x_1+x_2)(\xi_5^{'0}-\xi_5^0)\,.
\end{eqnarray}
The parameters in the expressions are as follows:
\begin{eqnarray}
\phi_3^0&=&\phi_6^0=f_X,\hspace{2.8cm}\psi_4^0=\psi_5^0=\frac12(f_X-\lambda_1)\,,\nonumber\\
\phi_4^0&=&\phi_5^0=\frac12(f_X+\lambda_1),\hspace{1.3cm}\phi_3'^0=\phi_6'^0=-\xi_5^0=\frac16(4\lambda_3-\lambda_2)\,,\nonumber\\
\phi_4'^0&=&\xi_4^0=\frac16(8\lambda_3-3\lambda_2),\hspace{1.2cm}\phi_5'^0=-\xi_5'^0=\frac16\lambda_2\,,\nonumber\\
\xi_4'^0&=&\frac16(12\lambda_3-5\lambda_2)\,.
\end{eqnarray}
The nonperturbative parameters $f_X$, $\lambda_1$ and $\lambda_2$ are determined with QCD sum rules to be
\begin{eqnarray}
f_{\Sigma}&=&(9.4\pm0.4)\times10^{-3}\; \mbox{GeV}^2,\hspace{0.8cm}\lambda_1=-(2.5\pm0.1)\times10^{-2}\; \mbox{GeV}^2,\nonumber\\
\lambda_2&=&(4.4\pm0.1)\times10^{-2}\; \mbox{GeV}^2,\hspace{0.8cm}\lambda_3=(2.0\pm0.1)\times10^{-2}\; \mbox{GeV}^2\label{sigmapara}
\end{eqnarray}
for $\Sigma$ and
\begin{eqnarray}
f_{\Xi}&=&(9.9\pm0.4)\times10^{-3}\; \mbox{GeV}^2,\hspace{0.8cm}\lambda_1=-(2.8\pm0.1)\times10^{-2}\; \mbox{GeV}^2,\nonumber\\
\lambda_2&=&(5.2\pm0.2)\times10^{-2}\; \mbox{GeV}^2,\hspace{0.8cm}\lambda_3=(1.7\pm0.1)\times10^{-2}\; \mbox{GeV}^2\label{sigmapara}
\end{eqnarray}
for $\Xi$.

%%%%%%%%%%%%%%%%%%%%%%%%%%%%%%%%%%%%%%%%%%%%%%%%%%%%%%%%%%%%%%%%%%%%%%%%%%%%%

\end{document}